%% file: paper-sigconf.tex
\documentclass[sigconf]{acmart}

\renewcommand\footnotetextcopyrightpermission[1]{} 
\usepackage{booktabs} 

\usepackage{algorithm}
\usepackage[noend]{algpseudocode}

\usepackage{fontenc}
\usepackage[utf8]{inputenc}
\usepackage{mathtools}

\usepackage{thmtools, thm-restate}
\usepackage{hyperref}
\usepackage{amsthm}

\begin{document}
\title{Learning to Customize Network Security Rules}

\author{Michael Bargury, Roy Levin, Royi Ronen}
\affiliation{%
	\institution{Microsoft, Israel}
}
\email{\string{t-mibarg, rolevin, royir\string}@microsoft.com}

\begin{abstract}
Security is a major concern for organizations who wish to leverage cloud computing. 
In order to reduce security vulnerabilities, public cloud providers offer firewall functionalities.
When properly configured, a firewall protects cloud networks from cyber-attacks.
However, proper firewall configuration requires intimate knowledge of the protected system, high expertise and on-going maintenance.

As a result, many organizations do not use firewalls effectively, leaving their cloud resources vulnerable.
In this paper, we present a novel supervised learning method, and prototype, which compute recommendations for firewall rules. 
Recommendations are based on sampled network traffic meta-data (NetFlow) collected from a public cloud provider.
Labels are extracted from firewall configurations deemed to be authored by experts. 
NetFlow is collected from network routers, avoiding expensive collection from cloud VMs, as well as relieving privacy concerns.

The proposed method captures network routines and dependencies between resources and firewall configuration. The method predicts IPs to be allowed by the firewall. 
A grouping algorithm is subsequently used to generate a manageable number of IP ranges. Each range is a parameter for a firewall rule.

We present results of experiments on real data, showing ROC AUC of 0.92, compared to 0.58 for an unsupervised baseline.
The results prove the hypothesis that firewall rules can be automatically generated based on router data, and that an automated method can be effective in blocking a high percentage of malicious traffic.

\end{abstract}

\newenvironment{problem}[1]{%
	\par\noindent%
	\renewcommand{\arraystretch}{1.2}%
	\tabularx{\textwidth}{|>{\bfseries}lX|c}%
	\hline%
	\multicolumn{2}{|l|}{%
		\raisebox{-\fboxsep}{\textsc{\Large #1}}%
	} \\[2\fboxsep]%
}{%
	\\ \hline%
	\endtabularx%
	\par%
}

\maketitle

\input{paperbody-conf}

\bibliographystyle{ACM-Reference-Format}
\bibliography{sigproc} 

\appendix
\newpage

\section{Appendix}

\begin{table}[h]
	\caption {Feature List} \label{tab:features} 
	\begin{tabular}{l}
		\hline \hline
		Distribution of communication over time \\
		\hline
		\% of inactive hours \\
		max over average of hourly \# of packets \\
		average of hourly \# of packets \\
		std of hourly \# of packets \\
		max over average of daily \# of packets \\
		average of daily \# of packets \\
		std of daily \# of packets \\
		\hline \hline
		Type of traffic \\
		\hline
		\% of TCP packets out of all packets sent \\
		\% of SYN TCP packets out of all packets sent \\
		\% of RESET TCP packets out of all packets sent \\
		\% of FIN TCP packets out of all packets sent \\
		\% of TCP packets out of all packets received \\
		\% of SYN TCP packets out of all packets received \\
		\% of RESET TCP packets out of all packets received \\ 
		\% of FIN TCP packets out of all packets received \\
		\hline \hline
		Breadth of communication \\
		\hline
		\% of VMs the IP communicates with  \\ 
		\# VMs the IP communicates with / \# IP packets observed \\
		\% of ports the IP uses, out of \# of possible ports \\
		\% of ports the IP communicates to out of \# of possible ports \\
		\# ports the IP uses / \# IP packets observed \\
		\# ports the IP communicates with / \# IP packets observed \\
		\hline
	\end{tabular}
	
	\phantom. A list of extracted features. Each feature is computed at four different levels, as described in Section \ref{feature-ext}. 
\end{table}

\appendix

\end{document}

%% file: paperbody-conf.tex
\section{Introduction}
Cloud security introduces unique opportunities as well as challenges, and is the top concern for organizations who wish to leverage the public cloud for their core business infrastructure~\cite{so2011cloud, idc2009survey}.
While most security breaches can be prevented by proper configuration~\cite{eshete2011early, khalil2013security}, the need for highly-customized configurations makes it prohibitively expensive for many organizations to remain protected over time~\cite{dourish2004security}. 
This is of increasing importance as large-scale computing is commoditized by the cloud, enabling organization to deploy advanced architectures.

\begin{figure}
	\caption{Recommendation Scenario}
	\centering
	\includegraphics[width=0.8\linewidth]{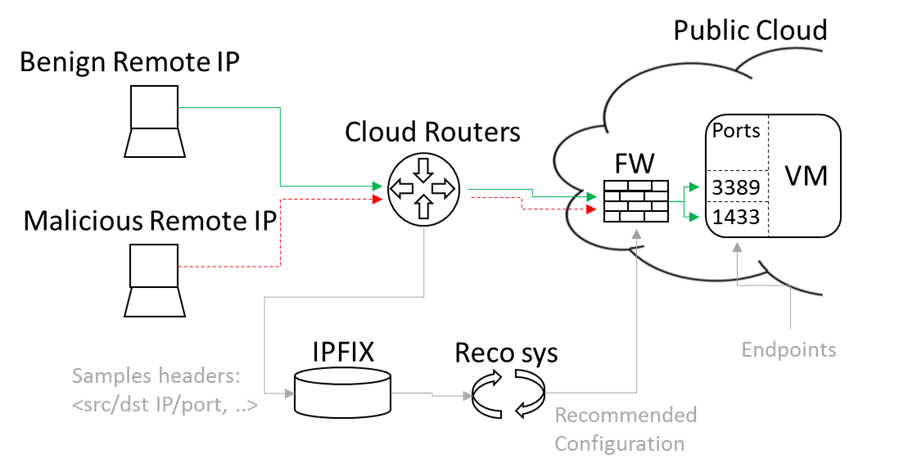}
	\label{fig:recosys}
	
	Benign and malicious IPs attempt to communicate with a cloud endpoint. 
	The recommender learns malicious and benign patterns, and generates a proper configuration.
\end{figure}

A network {\itshape endpoint} which consists of a Virtual Machine (VM) and an open port, is the gateway to the organization's virtual network (VNet).
Endpoints allow the VNet to communicate with other resources and users.
Unfortunately, endpoints might allow malicious intenders to gain access and compromise network assets.
In response, cloud providers allow control over endpoint access, using a firewall. The firewall has a list of allowed IP addresses and protocols.
Yet, by default, most public cloud providers allow endpoints be accessed from any IP~\cite{nsgdocs2016, awsdocs2016}.
Most of these endpoints remain with default configuration, because creating and managing white-lists is difficult: IP addresses change, and a misconfiguration can result in a broken service. Figure \ref{fig:recosys} shows an illustration of the recommendation scenario.

Automatic configuration and maintenance of firewall rules has been the subject of extensive research. Existing methods focus on capturing routine network usage with unsupervised methods, such as association rule mining applied on firewall logs\cite{saboori2010automatic}, and network traffic logs\cite{golnabi2006analysis,al2003firewall}. Capturing routine usage is also addressed in \cite{xu2012algorithms,molloy2012generative}, where formulation and methods are proposed in order to find models which reflect network state.
In contrast to our approach, previous work produces firewall rules from data of a single resource (losing dependencies between resources).
To the best of our knowledge, this work is the first to introduce a supervised approach to firewall rules generation. 
This enables us to automatically filter out malicious intenders and scanners which might be allowed by previous methods.
Also, we use traffic from network routers, which enables us to produce recommendations without the need to collect firewall logs from each machine.

Recommender systems have become very successful at predicting user-item interactions~\cite{ricci2011introduction}.
Despite their significance, few recommendation systems have been proposed for cyber-security.
Automated recommendations reduce the expertise level and maintenance costs required to manage security, while minimizing the chances of system interruption.
One such example is \cite{lyons2014recommender}, where the suggested system provides a list of actions to improve network security, including firewall updates. However, the system does not generate a full firewall configuration from scratch.

In this paper, we present a novel cyber-security recommendation system which generates a list of IP ranges as recommendations for an access white-list for network endpoint firewalls.
The system learns from existing white-list configuration authored by cyber-security experts, and predicts the IP ranges to be allowed for unconfigured endpoints.

\section{Algorithm}

\subsection{Dataset}\label{dataset}
Our dataset is sampled network traffic (NetFlow) from a cloud provider's routers in IPFIX format~\cite{claise2008specification}, collected over three weeks. The sampling ratio is one every four thousand packets.
Each raw sample represents a traffic {\itshape flow} which consists of: timestamp, source and destination IPs and ports, direction, protocol and TCP flags.
We refer to IPs that communicate with an endpoint as {\itshape remote IPs}.
Each sample describes communication between an endpoint and a remote IP. The processing of flow attributes into sample features is discussed in Section \ref{feature-ext}.
In total, we have a little over 10 million samples corresponding to 250TB of data.

Endpoints are matched to their firewall configuration. We observe that about 5\% of endpoints are manually configured, overriding the default (allowing all communication).
We assume that these endpoints are configured by domain experts. This assumption was verified with a sample of authors.
Samples corresponding to these endpoints are considered labeled data, where a sample is labeled positive if the remote IP is allowed and negative otherwise.

\subsection{Feature Extraction}\label{feature-ext}
Our model aims to distinguish between benign and malicious network traffic, learning from the labeled traffic described in Section \ref{dataset} (5\% of the entire dataset).
To capture the characteristics of a remote IP's behavior, we use features that fall into three categories:
(1) The {\itshape distribution} of communication over time, which consists of the percentage of inactive hours, average, maximum and standard deviation of daily and hourly communication volume;
(2) The {\itshape type} of traffic, which consists of percentage of incoming and outgoing TCP flags (Syn, Reset, Fin). These flags are sent when a communication flow has started or ended. Hence, they provide a good indicator to the number of new connections established;
(3) The {\itshape breadth} of communication, which consists of: number of VMs communicating with the remote IP, and the number of ports used.
We also use a combination of these features by applying a quadratic polynomial kernel. Table \ref{tab:features} provides a full feature list.

For every remote IP, each feature is computed at four different levels: (1) entire cloud; (2) organization; (3) VM; and (4) endpoint. The different levels allow the model to learn complex dependencies between the communication routine of a remote IP with an organization's environment, and its interactions with the cloud.

For intuition on the importance of using different levels, consider the breadth category.
High feature values imply that the remote IP is communicating with many different VMs, using various ports and different protocols.
For feature levels 2-4, this means the remote IP is strongly related to an organization's deployment, thus provides a good indicator that it should be allowed.
In contrast, for level 1 it means the remote IP communicates with many cloud deployments and organizations, which is a good indicator of automatic scanning that should be denied.

\begin{algorithm}
	\caption{IP Grouping. \\Sub procedures are described in Sub Procedure \ref{ipgroupingsub}}\label{ipgroup}
	\begin{algorithmic}[1]
		\State \textbf{input 1:} $p_1, \ldots, p_N \text{ - authorized IPs}$
		\State \textbf{input 2:} $L \text{ - max number of prefix sets}$
		\State \textbf{input 3:} $S \text{ - max prefix set size}$
		\State \textbf{output:}  $\text{a cover which satisfies definition }$\ref{cover-definition}
		\Procedure{FindMinCover}{$p_1, \ldots, p_N, L, S$}
		\State $A_{0,j} \gets 0$ $\forall 0\leq j \leq L$
		\State $A_{i,0} \gets \infty$ $ \forall 1\leq i \leq N$
		\For{$i \gets 1$ {\bfseries to} $N$}
		\For{$j \gets 1$ {\bfseries to} $L$}
		\State $s \gets arg\min_{k \leq i}{s.t.}$
		\State  \hskip\algorithmicindent \phantom. $|shared\text{ }prefix\text{ }set\text{ }of\text{ }p_k,p_i| \leq S$
		\State $A_{i,j} \gets \infty$
		\For{$k \gets s$ {\bfseries to} $i$}
		\State $I \gets shared\text{ }prefix\text{ }set\text{ }of\text{ }p_k,p_i$
		\State $x \gets noise(I)$
		\If{$A_{i,j} > x+A_{k-1,j-1}$}
		\State\Comment{update for i points, j prefix sets}
		\State $A_{i,j} \gets x+A_{k-1,j-1}$\Comment{min noise}
		\State $B_{i,j} \gets k$\Comment{first authorized IP in I}
		\State $C_{i,j} \gets I$
		\EndIf
		\EndFor
		\EndFor
		\EndFor
		\State $PrintCover(A,B,C)$ \Comment{iterate backwards to print results}
		\EndProcedure
	\end{algorithmic}
\end{algorithm}

\subsection{Learning Models}\label{learning-models}
The cost of denying a remote IP due to misclassifying can be high. If a remote IP communicates heavily with an endpoint, misclassification can break a service.
Therefore, a key modeling challenge is weighting misclassification.

To overcome this challenge we define a remote IP's {\itshape importance}, with respect to an endpoint, as the relative ratio of its communication with the endpoint.
We weight each sample by the remote IP's importance, forcing the model to focus its attention on high importance remote IPs.

We observe that the negative labeled samples (IPs to deny) represent 3\% of labeled samples described in Section \ref{dataset}, and only 5\% of the total number of traffic flows accounted for by the labeled samples.
This is due to the noisy nature of IP communication, enhanced by the exposure of the public cloud to automatic scanners.
Scanners may be legitimate, like crawlers, or malicious intenders such as vulnerability scanners.
To compensate for the very skewed nature of our classes, we normalize the weights of samples in each class by the total class weight.

We devise two recommendation methods, the first is used as a baseline.

\subsubsection{Baseline}\label{class-base}
The model collects a list of all remote IPs that communicate with a given endpoint in the three weeks learning period (the entire dataset), and predicts "allow" if and only if an IP is in that list. Due to the underlying sampling of our dataset described in Section \ref{dataset}, it is most likely that these IPs communicated with the endpoint many times. This model is unlikely to block routine communication, but it is prone to allow malicious intenders.

\subsubsection{SVM Classifier}\label{class-svm}
We use a binary Support Vector Machine (SVM) with a quadratic polynomial kernel to learn a weight vector for our features, described in Section \ref{feature-ext}, and predict our labels, described in Section \ref{dataset}. The classifier is then applied on each instance consisting of an endpoint and remote IP, to predict whether or not the IP should be allowed access to the endpoint.

\section{IP Grouping}
When producing a white-list, ranges are easier to comprehend and maintain than single IPs. Furthermore, ranges provide more generalization which improves our recommendation by allowing small fluctuations in the IP address.
We therefore devise a grouping algorithm using a dynamic programming approach, which transforms a list of IPs into CIDR formated~\cite{fuller2006classless} IP ranges.
The CIDR format is comprised of a base IP address and a number of least significant bits, that can vary within the range. An IP is covered by the range if by neglecting the specified number of least significant bits, it is the same as the base IP.
The goal of our algorithm is to find ranges which cover all IPs classified to be allowed, whilst keeping the number and size of these ranges as small as possible.
Running the algorithm multiple times with different constraints results in a configurable recommendation, allowing an organization to decide whether to emphasize reduced attack surface, or manageability and generalization.
Section \ref{exper} describes an experiment using the algorithm on real data. 

\floatname{algorithm}{Sub Procedure}
\begin{algorithm}
	\caption{IP Grouping}\label{ipgroupingsub}
	\begin{algorithmic}[1]
		\Procedure{PrintCover}{$A,B,C$}
		\If{$A_{N,L} = \infty$}
		\State $failed$ \Comment{impossible constraints}
		\Else
		\State $i \gets N$
		\State $j \gets L$
		\While{$i > 0$}
		\State $\text{print } C_{i,j}$
		\State $i \gets B_{i,j}-1$
		\State $j \gets j-1$
		\EndWhile
		\EndIf
		\EndProcedure
	\end{algorithmic}
\end{algorithm}
\begin{figure}
	\caption{IP Grouping Example for 3-Bit IP Addresses}
	\centering
	\includegraphics[width=0.9\linewidth]{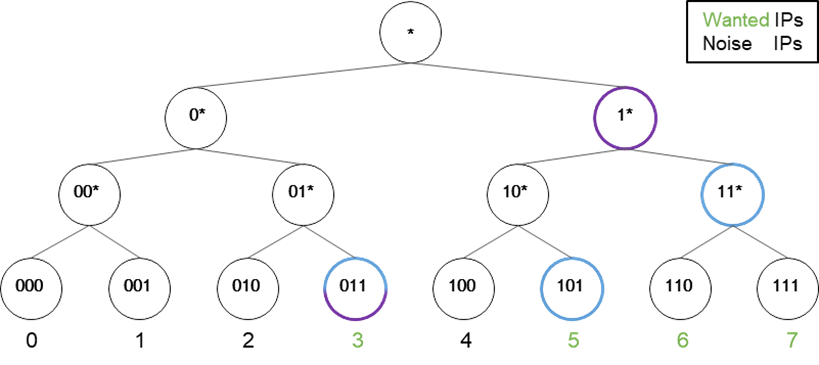}
	\label{fig:algo}
	
	Possible recommendations for $W=\{011, 101, 110, 111\}$. A cover with zero noise is presented in blue, and a best possible cover by two prefix sets is presented in purple.
\end{figure}

\subsection{Problem Statement}
Let $D$ be the set of natural numbers from $0$ to $2^{32}$ in their binary representation. $D$ represents all possible V4 IP addresses.
We define {\itshape authorized} IPs as those classified to be white-listed. Denote $D \supset W = \{p_1, \ldots, p_N\}$ the subset of all $N$ authorized IPs, and $p_i \in W$ the i'th authorized IP. 

Given a bit string of length $\leq 32$ denoted $d$, we define a {\itshape prefix set} $I \subset D$ to be the set of IPs which share $d$ as their most significant bit prefix. We say that $I$ is {\itshape generated} from $d$.
The {\itshape noise} of $I$ is $|I \setminus W|$, which corresponds to the number of unauthorized IPs that share $d$ as their significant bit prefix.
The {\itshape prefix notation} of a $I$ is its bit prefix $d$, followed by * to account for bits that can vary within the set.
Given two prefix sets, their {\itshape shared prefix} is the longest shared substring of their most significant bit prefixes. Their {\itshape shared prefix set} is the prefix set generated from their shared prefix.
For example, consider a 5-bit version of $D$. Let $W=\{11000,11001\}$.
Let $I_1$ be the prefix set of IP addresses with the prefix "110", and $I_2$ be the prefix set of IP addresses with the prefix "111".
The prefix notation of $I_1$ is "110*", its size is 4 and its noise is 2. The shared prefix of $I_1,I_2$ is "11", and their shared prefix set is the prefix set of IP addresses with the prefix "11".

Finally, we define a {\itshape cover} $C$ to be a set of pairwise disjoint $m$ prefix sets $I_i$, i.e $C=\{I_1, \dots, I_m\}$, which satisfies $W \subset \bigcupdot\limits_{i=1}^{m} I_{i}$.
We define it's noise as $\sum\limits_{i=1}^{m}|I_i \setminus W|$.

\begin{definition}\label{cover-definition}
	Given $W = \{p_1, \ldots, p_N\}$ authorized IPs, $L$ maximum number of prefix sets,
	and $S$ maximum prefix set size, our {\itshape objective} is to find a cover $C$ which satisfies:
\begin{alignat*}{2}
	\text{minimize }   & noise(C) \\
	\text{subject to } & W \subset \bigcupdot\limits_{i=1}^{m} I_{i} &,\ & \forall I_i \in C \\
					   & |I_i| \leq S &,\ & \forall 1\leq i\leq m\\
					   & m \leq L &\ & 
\end{alignat*}
\end{definition}

\subsection{The Grouping Algorithm}
The IP Grouping method (Algorithm \ref{ipgroup}) uses dynamic programming to find the solution to objective function in Definition \ref{cover-definition}.
The key to the algorithm is the following: after each iteration $i,j$, $A_{i,j}$ holds the minimum noise to cover at least the first $i$ authorized IPs, by cover length $\leq j$, with prefix set sizes $\leq S$.
For each $i,j$, the {\itshape FindMinCover} procedure in Algorithm \ref{ipgroup} iterates trough the different possibilities of adding the $i$'th IP to an existing cover of $j-1$ prefix sets and $i-1$ IPs. Hence, by assuming that the smaller covers that were computed in previous iterations are minimal, we are guaranteed to find the minimum cover in each new iteration. 
Furthermore, after the last iteration we have a minimum cover for all $N$ authorized IPs, which satisfies our objective function.

The run time complexity of the algorithm is O($N \times L \times S$). In practice, the number of prefix sets is limited by the number of rules allowed in the firewall which is typically small~\cite{nsgdocs2016, awsdocs2016}. Furthermore, the maximum prefix set size is also kept small to avoid large ranges (see Section \ref{exper}). Hence, in practice the run time complexity of the algorithm is O($N$).

\subsection{A Short Example}
Consider a simplified case of a 3-bit version of $D$:\\ $D = \{000,001,010,011,100,101,110,111\}$. \\Let $W=\{011, 101, 110, 111\}$. For simplicity, let $L=4$ and $S=8$.
Lines 2-19 of $FindMinCover$ (Algorithm \ref{ipgroup}) produce the matrices:
\[
A=
\begin{bmatrix}
0      & 0 & 0 & 0 & 0 \\
\infty & 0 & 0 & 0 & 0 \\
\infty & 6 & 0 & 0 & 0 \\
\infty & 5 & 2 & 0 & 0 \\
\infty & 4 & 1 & 0 & 0
\end{bmatrix}
,B,C=
\begin{bmatrix}
. & .   & .   & .   & .   \\
. & 1,011 & 1,011 & 1,011 & 1,011 \\
. & 1,*   & 2,101 & 2,101 & 2,101 \\
. & 1,*   & 2,1*  & 3,110 & 3,110 \\
. & 1,*   & 2,1*  & 3,11* & 3,11* 
\end{bmatrix}
\]
Using procedure $PrintCover$ of Sub Procedure \ref{ipgroupingsub} with these matrices, we get a minimum cover $\{11*,101,011\}$ with $noise=0$.
By initializing $j$ to the maximum number of prefix sets wanted $L' \leq L$ instead of $L$, one can get every minimum cover for $L' \leq L$. By summing over the entries of $A$, one can get the $noise$ of that cover. For example, setting $i \gets 2$ we get a minimum cover $\{1*,011\}$ with $noise=1$.
Figure \ref{fig:algo} shows these covers as leafs of a binary tree.
The reader will notice that the number of rules has become manageable.

\begin{figure}
	\centering
	\caption{ROC Curve}
	\includegraphics[width=0.75\linewidth,height=1.8in]{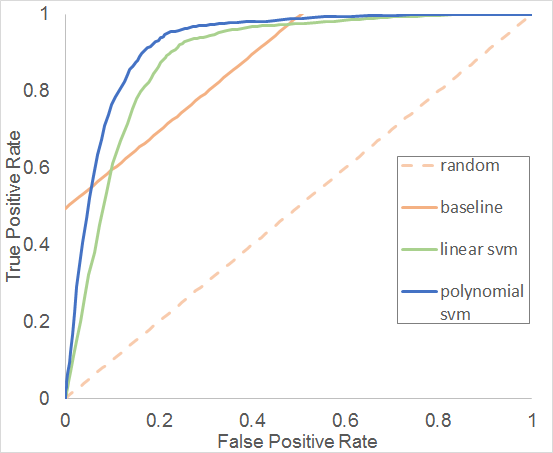}

	\label{fig:roc}
	
	Liner SVM, Polynomial SVM and the Baseline ROC curves.
\end{figure}

\section{Experiments and evaluation}\label{exper}
We start by comparing the results of our baseline and SVM classifier described in Section \ref{learning-models}, based on hypothesis validation. 
We use 60\% of our labeled data set for learning, 20\% for parameter search of the SVM regularization term and 20\% as a holdout set for final evaluation. Note that these percentages are as a part of the labeled data set described in Section \ref{dataset}.
For each remote IP and endpoint in the holdout set, the trained model predicts whether a domain expert would have included the remote IP in the endpoint's white-list. We then compare the prediction with the known firewall configuration.

\begin{figure}
	\caption{Classes separated by the model}
	\centering
	\includegraphics[width=0.95\linewidth]{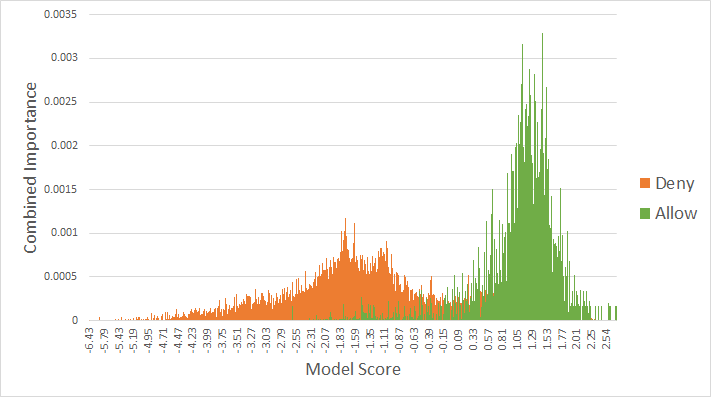}
	\label{fig:auc}
	
	Test set class separation, provided by the polynomial SVM.
\end{figure}

To measure the performance, we use AUC over a weighted ROC curve. 
The samples are weighted as described in Section \ref{learning-models}, and result in a balance between the classes of IPs that should be allowed and denied.
The firewall generation task requires both a high true positive rate which correspond to allowing access to authorized IPs, and a high true negative rate which correspond to denying access from unauthorized IPs.
Hence, the AUC metric over a ROC curve is a good measure for the success of our model.
We measure an AUC of 0.92 for our polynomial SVM model, compared to 0.58 for our baseline model.
Hence, our model provides an improvement of 0.34 in the AUC metric.
The improved AUC is a result of learning the features of white-listed remote IPs as opposed to allowing access to remote IPs that historically accessed the endpoint.
Figure \ref{fig:roc} shows an ROC curve comparison of the baseline model, a linear SVM and a polynomial SVM. 
Figure \ref{fig:auc} depicts the separation of classes in the test set, provided by the score of the polynomial SVM.

\begin{figure}[h!]
	\centering
	\caption{Feature Level Importance}
	\includegraphics[width=0.95\linewidth]{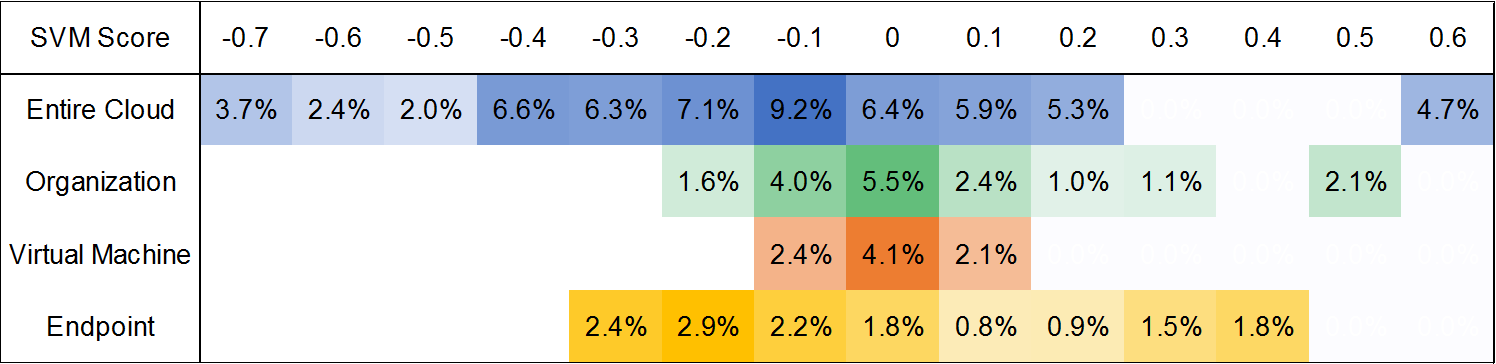}
	\label{fig:featureimportance}
	
	Cells represent the percentage of weight a feature level contributes to the final prediction.
\end{figure}

For a deeper understanding of the underlying behavior of our model, we analyze our feature importance. Figure \ref{fig:featureimportance} provides insights into the combined importance of features in each level, described at the end of Section \ref{feature-ext}.
We observe that features in the entire cloud level are the dominant ones for denying access, while allowing access can be attributed to a combination of entire cloud, organization and endpoint level features.

Algorithm \ref{ipgroup} is applied to the results of the model, to generate the final recommendation. We set $L=200$, the default number of firewall rules allowed by major public cloud providers ~\cite{nsgdocs2016, awsdocs2016}. In addition, we set $S=4096$, which can cover the IP range of a small Internet Service Provider ~\cite{fuller2006classless}, therefore can almost surly cover organizational network.

\section{Conclusions and Future Work}
In this work, we introduced a novel supervised method which recommends firewall rules for cloud endpoints.
Our method combines learning from firewall configurations created by domain experts as well as network traffic analysis to generate a recommendation which excludes malicious intenders.
Our experiments show that allowing all remote IPs that frequently interact with the endpoint is not likely to provide a good estimation of a firewall configured by a domain expert.
This stresses the importance of using new methods for firewall rule generation for cloud endpoints.
In future research, we plan to adapt the method suggested in this work to produce recommendations for other security and maintenance issues such as VPN configuration and VNET architecture.